\documentclass[12 pt,a4]{article}
\usepackage{amsmath,graphicx}
\input epsf
\begin{document}
\begin{flushright}
ITEP-TH-54/05
\end{flushright}
\vskip 1 cm
\begin{center}
\Large {\bf Monopole decay in the external electric field}
\end{center}
\vskip 2cm
\begin{center}
A.K. Monin
\end{center}
\vspace{3mm}
\begin{center}\emph{Institute of Theoretical and Experimental
Physics, \\ B.Cheremushkinskaya 25, Moscow, Russia}
\end{center}
\vspace{3mm}
\begin{center}
\emph{Moscow State University, Physics Department, \\ Vorobyevy
Gory 1, Moscow, Russia}
\end{center}
\begin{center}
e-mail: monin@itep.ru
\end{center}
\vspace{5mm}
\vskip1.3cm
\begin{abstract} The possibility of the magnetic
monopole decay in the constant electric field is investigated and
the exponential factor in the probability is obtained. Corrections
due to Coulomb interaction are calculated. The relation between
masses of particles for the process to exist is obtained.
\end{abstract}
\newpage
\section{Introduction}
Tunneling processes are very interesting nonperturbative
phenomena. One can find the example of such process already in
quantum mechanics, where it causes the energy splitting. There are
tunneling processes in the field theory as well, for instance,
pair production in external electromagnetic fields \cite{sch},
decay of the false vacuum \cite{kobz vol okun, Coleman2}.

In some spontaneously broken gauge theories there are magnetic
monopole and dyon solutions. It is supposed that they can be
produced in strong enough external electromagnetic fields. In the
weak coupling regime their masses are huge, and sizes are of order
$\sim \frac{1}{e^{2}M}$. The probability of magnetic monopole pair
production in constant magnetic field was calculated in the work
of Affleck and Manton \cite{aff1} using the instanton method. In
the work of Bachas and Porrati \cite{bachporr} the rate of pair
production of open bosonic and supersymmetric strings in a
constant electric field was calculated exactly. In the work of
Gorsky, Saraikin and Selivanov \cite{gor} stringy deformed
probability of monopole and W-boson pair production was obtained
quasiclassically. It is possible for particles like monopole, dyon
or W-boson to decay nonperturbatively in external fields
\cite{gor}. Monopole can also decay nonperturbatively in the
external 2-form field \cite{Gor Sel junct}.

In this paper we consider the process of magnetic monopole decay
into electron and dyon, and W-boson decay into dyon and monopole
using instanton method. Euclidean configuration corresponding to
the monopole decay is represented on the fig.\ref{classical
traject}. Exponential factor in the probability is given by the
minimum of the electron, dyon and monopole total effective action
(see (\ref{eff action})). This leads to (\ref{classical action})
for the classical action. When monopole mass is equal to zero, one
gets well-known result for exponential dependence of probability
for pair production in external field. Dyon is not pointlike
particle, so to apply this method for calculation one must imply
that the size of dyon is much smaller then the size of
electron-dyon loop. So, there is some condition imposed on the
external field. The approximation used in this case is analogous
to the thin wall approximation in the problem of the false vacuum
decay. There is also condition of the dyon stability
$M_{d}<M_{m}+m$, where $m$, $M_{d}$, $M_{m}$ are masses of
electron, dyon and monopole respectively. Contrary to spontaneous
pair production, the process of the particle decay doesn't occur
for arbitrary masses. It is shown that for the case when the
relation (\ref{condition on masses}) is fulfilled there are two
negative eigenmodes, so, there is no decay at all. Coulomb
corrections are taken into account similar to the work \cite{aff2}
in the limit $M_{d}\gg m$.
\section{Spontaneous particle production}
\label{spont prod}
\subsection{$e^{-}e^{+}$ pair production} One can obtain
Schwingers \cite{sch} result for probability of $e^{-}e^{+}$ pair
production summing the diagrams for vacuum amplitude similar to
one represented on fig.\ref{diagram}:
\begin{figure}[h]
\centerline{\epsfbox{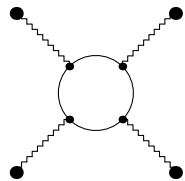}}\caption{Diagrams for pair production
process} \label{diagram}
\end{figure}
\begin{equation}
S_{0}=\langle0\mid S\mid 0\rangle=det(i\widehat{\partial}-m)\cdot
det\left
(1-\frac{i}{i\widehat{\partial}-m}(-ie\widehat{A})\right).
\end{equation}
The probability connected with the amplitude is of the form
\begin{equation}
|S_{0}|^{2}=\exp( -\int d^{4}x w(x)),
\end{equation}
where $w(x)$ is the probability of pair production per unit time
per unit volume.
\begin{equation}
2\ln S_{0}=sp\ln \left
(((\widehat{P}-e\widehat{A}(x))^{2}-m^{2})\frac{1}{P^{2}-m^{2}}\right).
\end{equation}
Using the representation
\begin{equation}
\ln\frac{a}{b}=\int_{0}^{\infty}
\frac{ds}{s}(e^{is(b+i\varepsilon)}-e^{is(a+i\varepsilon)}),
\end{equation}
one obtains
\begin{equation}
\Gamma=w(x)=-\frac{1}{(2\pi)^{2}}\int_{0}^{\infty}\frac{ds}{s^{2}}\left
(eE\coth(eEs)-\frac{1}{s}\right) Re(ie^{-is(m^{2}-i\varepsilon)}).
\end{equation}
So, the probability for the process of $e^{-}e^{+}$ pair
production looks as follows
\begin{equation}
\Gamma = 2 \frac{(eE)^{2}}{(2
   \pi)^{3}}\sum^{\infty}_{n=1}\frac{1}{n^{2}}\exp\left(-\frac{\pi
   m^{2}}{eE}n\right)
\end{equation}
\subsection{Instanton method}
\begin{figure}[h] \centerline{\epsfbox{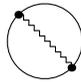}} \caption{Photon
exchange in the loop}\label{photon interchange}
\end{figure}

The result obtained in the previous section is valid only for
small coupling constant, since it doesn't take into account self
interaction of electron loop (see fig.\ref{photon interchange}).
Authors of \cite{aff1} have derived the expression for probability
of monopole-antimonopole pair creation in external magnetic field
using instanton method. As we know the rate of decay follows from
the imaginary part of the ground state energy
\begin{equation}
\Gamma'=2\mathrm{Im}E_{0},
\end{equation}
where the energy $E_{0}$ is obtained from
\begin{equation}
e^{-E_{0}T}\approx \lim_{T\rightarrow
\infty}\int\mathcal{D}x\mathrm{e}^{-S[x]}.
\end{equation}
Using the WKB approximation one can do the integral and find the
necessary probability per unit time per unit volume
\begin{equation}
\Gamma=2\frac{\mathrm{Im}E_{0}}{V}=2\mathrm{Im}\Delta\mathrm{e}^{-S_{cl}},
\end{equation}
where $S_{cl}$ is the classical action calculated on the instanton
solution, and $\Delta$ is one loop factor arising from the second
variation of action. We should note that operator of the second
variation must have one and only one negative eigenvalue. Affleck
and Manton have calculated $\Delta$ and $S_{cl}$ with the
following result
\begin{equation}
\Gamma_{M}=\frac{(gB)^{2}}{(2\pi)^{3}}\mathrm{e}^{-\frac{\pi
M^{2}}{gB}+\frac{g^{2}}{4}},
\end{equation}
where $M$, $g$ are the mass and the charge of the monopole and $B$
is the
strength of the external magnetic field. \\
The exponential factor in the probability can be immediately
obtained by minimizing the effective action
\begin{equation}
S_{eff}=ML-gBQ,
\end{equation}
where L is  the length of classical monopole trajectory in the
magnetic field, Q is the area restricted by this trajectory. Note
that recently instanton approach has been used for calculation of
the probability of the pair production in the nonhomogeneous
fields \cite{inhom fields} and in gravitational background
\cite{pioline troost}.
\section{Monopole decay}
\label{problem}
\subsection{Exponential factor in probability}
Now let us turn to the calculation of probability of the monopole
decay in the external electric field. It was argued \cite{gor}
that particles like monopole (W-boson) can decay in the external
electric (magnetic) field into electron and dyon (dyon and
antimonopole), the junction such as (DeM) naturally appears in the
string theory. To find the probability we have to calculate the
correction to Green's function in the presence of electron and
dyon. Green's function of free heavy monopole in external
electro-magnetic field in Euclidean time can be written as:
\begin{equation}
G(T,0;0,0)=\int\mathcal{D}y_{\mu}\mathrm{e}^{-M_{m}\int\sqrt{\dot{y_{\mu}}^{2}}dt}\sim
\mathrm{e}^{-M_{m}T}
\end{equation}
\begin{figure}[h]
\centerline{\epsfbox{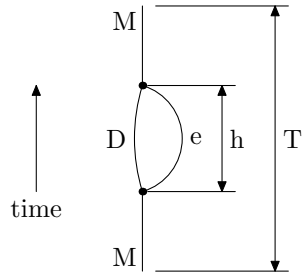}} \caption{Classical trajectories of
particles in Euclidean time} \label{classical traject}
\end{figure}
Taking into account one bounce correction we have
\begin{equation}
\begin{array}[c]{rl}
G(T,0;0,0)_{bounce}\sim&\int\mathcal{D}x\mathcal{D}z\exp\bigl(-m\int
\sqrt{\dot{x}^{2}}dt-M_{d}\int
\sqrt{\dot{z}^{2}}dt-M_{m}(T-h)\bigr.
\\ \\ -&ie\int (A_{\mu}(x)+A_{\mu}^{ext}(x))dx_{\mu}+ie\int
(A_{\mu}(z)+A_{\mu}^{ext}(z))dz_{\mu}
\\ \\  \bigl.-&\int\frac{1}{4}F_{\mu\nu}^{2}d^{4}x\bigr),
\end{array}
\end{equation}
where expression in the exponent is the well-known action for
relativistic particles interacting with electromagnetic field.
This correction is the first term of expansion of full Green's
function $\sim e^{-(M_{m}+\delta M_{m})T}$, and as we know, twice
the imaginary part of the mass yields the probability of the
decay. We consider the external field in Minkowski space as
constant electric field aligned in the $x_{1}$ direction. The
Euclidean version of the field is of the form
\begin{equation}
A_{\mu}=\frac{i}{2}\left(Ex^{1},-Ex^{0},0,0\right).
\end{equation}
Euclidean action for small coupling constant $e^{2}$ can be
written as
\begin{equation}
S=ml+M_{d}L-eEQ-M_{m}h, \label{eff action}
\end{equation}
where $l,L$ are the lengths of the trajectories and $Q$ is the
area of the region restricted by trajectories. We'll do path
integral by steepest descent approximation. At first we find
solution to the classical equation of motion. For the electron we
have
\begin{equation}
m\frac{d}{dt}\frac{\dot{x}_{\mu}}{\sqrt{\dot{x}^{2}}}=-ieF_{\mu\nu}\dot{x}_{\nu},
\end{equation}
while for the dyon
\begin{equation}
\frac{d}{dt}\frac{\dot{z}_{\mu}}{\sqrt{\dot{z}^{2}}}=ieF_{\mu\nu}\dot{z}_{\nu},
\end{equation}
where we neglect interaction between particles, i.e. Coulomb
effects. Solutions to these equations are arcs of circles with
radii $r=\frac{m}{eE}$, $R=\frac{M_{d}}{eE}$.
\begin{figure}[h]
\centerline{\epsfbox{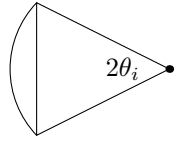}} \caption{Square of
segment}\label{segment}
\end{figure}
Introducing the angles that define the lengths of the arcs
$\theta_{1},\theta_{2}$ for electron and dyon respectively
(fig.\ref{segment}) we have the action:
\begin{equation}
\begin{array}[c]{rl}
S_{cl}=& 2mr\theta_{1}+2M_{d}R\theta_{2}-2M_{m}r\sin\theta_{1}- \\
\\ - & eE\theta_{1}r^{2}+\frac{1}{2}eEr^{2}\sin 2\theta_{1}
-eE\theta_{2}R^{2}+\frac{1}{2}eER^{2}\sin 2\theta_{2}.
\end{array}
\end{equation}
In order to find $\theta_{1},$ $\theta_{2}$, $r$, $R$ we minimize
action taking into account that
\begin{equation}
r\sin\theta_{1}=R\sin\theta_{2}.
\end{equation}
Adding to the action term $(r\sin\theta_{1}-R\sin\theta_{2})$ with
the Lagrange multiplier $\lambda$ one gets
\begin{equation} \left\{
\begin{array}[c]{rcl}
2mr-2M_{m}r\cos\theta_{1}+\lambda r\cos\theta_{1}-eEr^{2}+eEr^{2}\cos 2\theta_{1}&=&0, \\
2M_{d}R-\lambda R\cos\theta_{2}-eER^{2}+eER^{2}\cos 2\theta_{2}&=&0, \\
r\sin\theta_{1}-R\sin\theta_{2}&=&0, \\
2m\theta_{1}-2M_{m}\sin\theta_{1}+\lambda\sin\theta_{1}-2eE\theta_{1}r+eEr\sin
2\theta_{1}&=&0, \\
2M_{d}\theta_{1}-\lambda\sin\theta_{2}-2eE\theta_{2}R+eER\sin
2\theta_{2}&=&0.
\end{array}
\right.
\end{equation}
Solution to this system is
\begin{equation} \left\{
\begin{array}[c]{rcl}
r&=&\frac{m}{eE}, \\
R&=&\frac{M_{d}}{eE}, \\
\theta_{1}&=&\arccos\frac{M_{m}^{2}+m^{2}-M_{d}^{2}}{2mM_{m}}, \\
\theta_{2}&=&\arcsin\left(\frac{m}{M_{d}}\sin\theta_{1}\right),
\end{array}
\right. \label{rel btw ang}
\end{equation}
which amounts to the action
\begin{equation}
\begin{array}[c]{rl}
S_{cl}=&
\frac{m^{2}}{eE}\arccos\frac{M_{m}^{2}+m^{2}-M_{d}^{2}}{2mM_{m}}+
\frac{M_{d}^{2}}{eE}\arccos\frac{M_{m}^{2}-m^{2}+M_{d}^{2}}{2M_{d}M_{m}}- \\ \\
- & \frac{mM_{m}}{eE}\sqrt{1-\left(
\frac{M_{m}^{2}+m^{2}-M_{d}^{2}}{2mM_{m}}\right)^{2}}.
\end{array}
\label{classical action}
\end{equation}
This result, of course, could be obtained in the way similar to
that described in \cite{selvol} for the problem of induced false
vacuum decay. Classical trajectories of electron, dyon and
monopole are determined from the equilibrium condition in vertex
via mechanical analogy ($m, M_{d}$ are surface tensions and
$M_{m}$ external force)
\begin{equation} \left\{
\begin{array}[c]{c}
m\sin\theta_{1}=M_{d}\sin\theta_{2}, \\
M_{m}-m\cos\theta_{1}-M_{d}\cos\theta_{1}=0.
\end{array}
\right.
\end{equation}
It follows from these equations that if $M_{m}$ equals to zero
then $m$ and $M_{d}$ will be equal to each other, i.e. creation of
particles with different masses is accompanied by decay of some
massive particle. Also we should note that such action for dyon
and monopole implies that these particles have no size. However,
since these particles are not pointlike the approximation used
works only for small enough external field. The size of dyon
($\sim\frac{1}{e^{2}M_{d}}$) should be much smaller then the size
of electron-dyon loop ($\sim\frac{m}{eE}$), so the field should
obey the condition
\begin{equation}
E\ll emM_{d}.
\end{equation}
Also we should make some comment on the limit $M_{m}=0$, which
implies the condition $m=M_{d}$. Intuitively it seems to coincide
with the result of circular symmetrical case of pair production.
The action for this case, as one can see from (\ref{classical
action}), becomes
\begin{equation}
\frac{\pi m^{2}}{eE}.
\end{equation}
which is the same as in the circular symmetric case.
\subsection{W-boson decay}
Let us make a short digression on the W-boson decay in the
magnetic field. It was argued that W-boson can decay
nonperturbatively in external magnetic field into monopole and
dyon as magnetic monopole in external electric field \cite{gor}
into electron and dyon. However, since strong enough magnetic
fields occur in cosmology more often then strong electric fields
the formula for W-boson decay could be even more useful. The
problem of W-boson decay is identical to one of monopole decay
since there is dual symmetry between electric and magnetic fields.
So, the result for the probability of W-boson decay reads as
\begin{equation}
\Gamma_{w}=A\mathrm{e}^{\frac{M_{m}^{2}}{gB}\arccos\frac{M_{w}^{2}+M_{m}^{2}-M_{d}^{2}}{2M_{w}M_{m}}+
\frac{M_{d}^{2}}{gB}\arccos\frac{M_{w}^{2}-M_{m}^{2}+M_{d}^{2}}{2M_{d}M_{w}}-
\frac{M_{w}M_{m}}{gB}\sqrt{1-\left(
\frac{M_{w}^{2}+M_{m}^{2}-M_{d}^{2}}{2M_{w}M_{m}}\right)^{2}}},
\end{equation}
where $M_{w}$, $M_{m}$, $M_{d}$ are masses of W-boson, monopole
and dyon respectively, $B$ is external magnetic field and $g$ is
magnetic charge of monopole.
\subsection{Determinant}
Now we return to the monopole decay. It was mentioned before that
the classical trajectories of electron and dyon are the arcs of
the circles. For the electron we have
\begin{equation} \left\{
\begin{array}[c]{rcl}
x_{0}^{cl}&=&r\sin (2\theta_{1}t-\theta_{1}), \\
x_{1}^{cl}&=&-r\cos (2\theta_{1}t-\theta_{1})+r\cos\theta_{1}.
\end{array}
\right.
\end{equation}
Operator of the second variation looks as follows
\begin{equation}
-\frac{m}
{\sqrt{\dot{x}_{cl}^{2}}}\delta_{\mu\nu}\frac{d^{2}}{dt^{2}}+
\frac{m}{(\dot{x}_{cl}^{2})^{3/2}}\dot{x}^{cl}_{\mu}\dot{x}^{cl}_{\nu}\frac{d^{2}}{dt^{2}}+
\frac{m}{(\dot{x}_{cl}^{2})^{3/2}}(\ddot{x}^{cl}_{\mu}\dot{x}^{cl}_{\nu}+
\dot{x}^{cl}_{\mu}\ddot{x}^{cl}_{\nu})\frac{d^{2}}{dt^{2}}-
ieF_{\mu\nu}\frac{d}{dt},
\end{equation}
where
\begin{equation}
\sqrt{\dot{x}_{cl}^{2}}=2\theta_{1}r=2\theta_{1}\frac{m}{eE}.
\end{equation}
Let us define
\begin{equation}
a=2\theta_{1}t-\theta_{1},
\end{equation}
then we get the equations for eigenfunctions and eigenvalues
\begin{equation} \left\{
\begin{array}[c]{rcl}
-\frac{1}{4\theta_{1}}f_{0}''+\frac{1}{4\theta_{1}}\cos2af_{0}''-\sin2af_{0}'+\frac{1}{4\theta_{1}}\sin2af_{1}''
+\cos2af_{1}'-f_{1}'&=&\frac{\lambda}{eE} f_{0}, \\ \\
\frac{1}{4\theta_{1}}\sin2af_{0}''+\cos2af_{0}'+f_{0}'-\frac{1}{4\theta_{1}}f_{1}''
-\frac{1}{4\theta_{1}}\cos2af_{1}''+\sin2af_{1}'&=&\frac{\lambda}{eE}
f_{1}, \\ \\
-\frac{1}{2\theta_{1}}f_{2}''&=&\frac{\lambda}{eE}f_{2}, \\
\\-\frac{1}{2\theta_{1}}f_{3}''&=&\frac{\lambda}{eE} f_{3}.
\end{array} \right.
\label{equation second var}
\end{equation}
To solve these equations denote
\begin{equation}
\begin{array}[c]{rcl}
F=f_{0}+if_{1}, \\
\bar{F}=f_{0}-if_{1},
\end{array}
\end{equation}
and multiplying the second equation in (\ref{equation second var})
by $i$ and adding to the first equation we get
\begin{equation}
-\frac{1}{4\theta_{1}}F''+\frac{1}{4\theta_{1}}\mathrm{e}^{2ia}\bar{F}''+i\mathrm{e}^{2ia}\bar{F}'+i\bar{F}'
=\frac{\lambda}{eE} F.
\end{equation}
One can obtain complex conjugated equation by multiplying the
second equation by $i$ and subtracting it from the first one. The
shape of the equation tells us the form of the solution
\begin{equation} \left\{
\begin{array}[c]{rcl}
F&=&\mathrm{e}^{-ia}g, \\
g&=&g_{1}+ig_{2}.
\end{array}
\right.
\end{equation}
Upon this substitution we have
\begin{equation} \left\{
\begin{array}[c]{rcl}
-\frac{1}{2\theta_{1}}g_{2}''-2\theta_{1}g_{2}&=&\frac{\lambda}{eE}
g_{2}, \\ \lambda g_{1}&=&0,
\end{array}
\right.
\end{equation}
and
\begin{equation} \left\{
\begin{array}[c]{rcl}
f_{0}^{e}(t)&=&\left(A_{1}\cos\omega_{1} t+B_{1}\sin\omega_{1}
t\right)\sin(2\theta_{1}t-\theta_{1}), \\
f_{1}^{e}(t)&=&-\left(A_{1}\cos\omega_{1} t+B_{1}\sin\omega_{1}
t\right)\cos(2\theta_{1}t-\theta_{1}),
\end{array}
\right.
\end{equation}
\begin{equation}
\lambda=eE\left(\frac{\omega_{1}^{2}}{2\theta_{1}}-2\theta_{1}\right).
\label{lambda 1}
\end{equation}
Similar manipulations for the dyon yield
\begin{equation} \left\{
\begin{array}[c]{rcl}
f_{0}^{d}(t)&=& \left(A_{2}\cos\omega_{2} t+B_{2}\sin\omega_{2}
t\right)\sin(2\theta_{2}t-\theta_{2}), \\
f_{1}^{d}(t)&=&\left(A_{2}\cos\omega_{2} t+B_{2}\sin\omega_{2}
t\right)\cos(2\theta_{2}t-\theta_{2}).
\end{array}
\right.
\end{equation}
\begin{equation}
\lambda=eE\left(\frac{\omega_{2}^{2}}{2\theta_{2}}-2\theta_{2}\right).
\label{lambda 2}
\end{equation}
There are zero eigenmodes for every particle corresponding to the
translations. The mode corresponding to the translation in $x_{1}$
direction is
\begin{equation} \left\{
\begin{array}[c]{rcl}
f_{0}&=&0, \\
f_{1}&=&1.
\end{array}
\right.
\end{equation}
\begin{figure}[h]
\centerline{\epsfbox{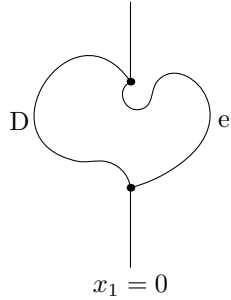}} \caption{The allowed perturbed
configuration} \label{allowed configuration}
\end{figure}
However, these perturbations for electron and dyon are not
independent. Since worldline of monopole is straight with
$x_{1}=0$ (see fig.(\ref{allowed configuration})), the unification
of distorted trajectories should be closed. The resulting
electron-dyon loop should have vertexes with at $x_{1}=0$. To
satisfy the last condition one can require
\begin{equation}
f_{1}(0)=f_{1}(1),
\end{equation}
and using zero mode translate perturbed trajectory to $x_{1}=0$.
To satisfy the former condition one must require
\begin{equation}
f^{e}_{0}(1)-f^{e}_{0}(0)=f^{d}_{0}(1)-f^{d}_{0}(0). \label{equal
time diff}
\end{equation}
Let us find solutions consistent with these restrictions. For the
solution with factor $\cos\omega t$ we get
\begin{equation}
f_{1}(0)=f_{1}(1)=A\cos\theta=A\cos\omega\cos\theta,
\end{equation}
so, $\omega=2 \pi n$. For the solution with factor $\sin\omega t$
we get
\begin{equation}
f_{1}(0)=f_{1}(1)=0=B\sin\omega\cos\theta,
\end{equation}
that is $\omega=\pi n$. Finally we have
\begin{equation} \left\{
\begin{array}[c]{rcl}
f_{0}^{e,d}(t)&=&B_{1,2}\sin\pi t\sin(2\theta_{1,2}t-\theta_{1,2}), \\
f_{1}^{e,d}(t)&=&\mp B_{1,2}\sin\pi
t\cos(2\theta_{1,2}t-\theta_{1,2}),
\end{array}
\right. \label{sin solution}
\end{equation}
with eigenvalues
\begin{equation}
\lambda_{n}^{e,d}=eE\left(\frac{(\pi
n)^{2}}{2\theta_{i}}-2\theta_{i}\right), \label{sin eigenvalue}
\end{equation}
and
\begin{equation} \left\{
\begin{array}[c]{rcl}
f_{0}^{e,d}(t)&=&A_{1,2}\cos 2\pi n t\sin(2\theta_{1,2}t-\theta_{1,2}), \\
f_{1}^{e,d}(t)&=&\mp A_{1,2}\cos 2\pi n
t\cos(2\theta_{1,2}t-\theta_{1,2}),
\end{array}
\right. \label{cos solution}
\end{equation}
with eigenvalues
\begin{equation}
\lambda_{n}^{e,d}=eE\left(\frac{(2 \pi
n)^{2}}{2\theta_{i}}-2\theta_{i}\right). \label{cos eigenvalue}
\end{equation}
It is obvious that first set of solutions already fulfills
(\ref{equal time diff}), so, for this set we have independent
perturbations for the electron and for the dyon. The second set do
not fulfills (\ref{equal time diff}) for arbitrary $A_{1}$ and
$A_{2}$. They must satisfy the condition
\begin{equation}
A_{1}\sin\theta_{1}=A_{2}\sin\theta_{2},
\end{equation}
thus
\begin{equation}
\frac{A_{1}}{A_{2}}=\frac{m}{M_{d}}.
\end{equation}
\begin{figure}[h]
\centerline{\epsfbox{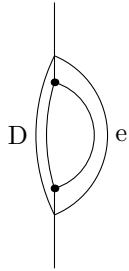}}\caption{Perturbation
corresponding to the increasing of radii of trajectories}
\label{radii increasing}
\end{figure}
It is quite difficult to find full quantitative expression for the
determinant, but we can investigate it qualitatively. We are
interested in the presence of negative eigenmodes. It is easy to
see from (\ref{cos eigenvalue}) that if $n=0$, some eigenvalue
becomes negative. Corresponding eigenfunctions are proportional to
the classical solutions
\begin{equation} \left\{
\begin{array}[c]{rcl}
f_{0}^{e}&=&a m\sin(2\theta_{1}t-\theta_{1}), \\
f_{1}^{e}&=&-a m\cos(2\theta_{1}t-\theta_{1}),
\end{array}
\right.
\end{equation}
\begin{equation} \left\{
\begin{array}[c]{rcl}
f_{0}^{d}&=& a M_{d}\sin(2\theta_{2}t-\theta_{2}), \\
f_{1}^{d}&=& a M_{d}\cos(2\theta_{2}t-\theta_{2}),
\end{array}
\right.
\end{equation}
and they are responsible for simultaneous increasing of the radii
of the trajectories (see fig.\ref{radii increasing}), where $a$ is
the corresponding parameter. Similar negative eigenmode exists in
the case of spontaneous pair production (see for example
\cite{aff2}). One can calculate the action on perturbed
configuration, expand it in power series by small deviation from
classical radius and find that the quadratic correction is
negative. Hence this perturbation is connected to the negative
eigenmode. Indeed, correction to the action proportional to
$a^{2}$ reads as
\begin{equation}
S_{(2)}=\frac{m^{2}a^{2}}{2}(-2\theta_{1}+\sin2\theta_{1})+
\frac{M_{d}^{2}a^{2}}{2}(-2\theta_{2}+\sin2\theta_{2})<0,
\end{equation}
since $x>\sin x$. \\
We can also consider the first set of solutions (\ref{sin
solution}). The solution corresponding to the particle with
smaller mass, i.e. to the electron, for $n=1$ is
\begin{equation} \left\{
\begin{array}[c]{rcl}
f_{0}^{e}&=&B_{1}\sin\pi t\sin(2\theta_{1}t-\theta_{1}), \\
f_{1}^{e}&=&-B_{1}\sin\pi t\cos(2\theta_{1}t-\theta_{1}),
\end{array}
\right. \label{specific eigenmode}
\end{equation}
and
\begin{equation}
\lambda^{e}=eE\left(\frac{\pi^{2}}{2\theta_{1}}-2\theta_{1}\right).
\end{equation}
\begin{figure}[h]
\centerline{\epsfbox{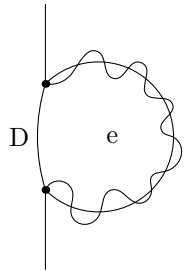}}\caption{Perturbation of
the electron's trajectory} \label{Pert of the electrons traject}
\end{figure}
It is obvious that $\lambda<0$ when $\theta_{1}>\frac{\pi}{2}$
(see fig.\ref{Pert of the electrons traject}) that is
\begin{equation}
M_{d}^{2}>M_{m}^{2}+m^{2}, \label{condition on masses}
\end{equation}
and the condition of dyon stability is
\begin{equation}
M_{d}^{2}<\left(M_{m}+m\right)^{2}=M_{m}^{2}+m^{2}+2mM_{d}.
\label{dyon stability}
\end{equation}
One can also calculate correction to the action
\begin{equation}
S_{(2)}=\left(-\frac{1}{2}\theta_{1}+\frac{\pi^{2}}{8\theta_{1}}\right)B_{1}^{2}.
\end{equation}
It is negative provided that condition on masses (\ref{condition
on masses}) is fulfilled. Since the presence of the negative
eigenmode connected to the changing of configuration's size do not
depend on this condition, we have two negative eigemodes in this
case. Hence, there is no decay at all provided that condition
(\ref{condition on masses}) is true. In this case the action on
the Euclidean configuration provides the nonperturbative
renormalization of monopole's mass.
\subsection{Coulomb corrections}
\begin{figure}[h]
\centerline{\epsfbox{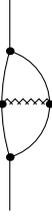}} \caption{Coulomb
interaction}\label{Coulomb}
\end{figure}
Now we'll calculate the quantum corrections due to photon exchange
between electron and dyon (see fig.\ref{Coulomb}). We know that
\begin{equation}
\int
\mathcal{D}AA_{\mu}(x)A_{\nu}(x^{'})\mathrm{e}^{-\frac{1}{4}\int
F_{\mu\nu}^{2}d^{4}x}=\frac{1}{4\pi^{2}}\frac{g_{\mu\nu}}{(x-x^{'})^{2}},
\end{equation}
and therefore using  the relation above we obtain the contribution
to the path integral from electromagnetic field
\begin{equation}
\begin{array}{c}
\int \mathcal{D}A\mathrm{e}^{-\frac{1}{4}\int
F_{\mu\nu}^{2}d^{4}x+ie\oint
A_{\mu}(x)dx_{\mu}}= \\
=\int \mathcal{D}A\mathrm{e}^{-\frac{1}{4}\int
F_{\mu\nu}^{2}d^{4}x}\left(1+ie\oint
A_{\mu}(x)dx_{\mu}+\frac{1}{2!}(ie)^{2}\oint
A_{\mu}(x)dx_{\mu}A_{\nu}(x^{'})dx_{\nu}^{'}+...\right)
\\ =\left(1-\frac{e^{2}}{8\pi^{2}}\oint\oint
\frac{g_{\mu\nu}}{(x-x^{'})^{2}}dx_{\mu}dx_{\nu}^{'}+...\right)=\mathrm{e}^{-\frac{e^{2}}{8\pi^{2}}\oint\oint
\frac{g_{\mu\nu}}{(x-x^{'})^{2}}dx_{\mu}dx_{\nu}^{'}}.
\end{array}
\end{equation}
We have to do the integral over the path which consists of
trajectories of particles, schematically
\begin{equation}
\oint\oint=\int_{electron}\int_{electron}+\int_{dyon}\int_{dyon}-
2\int_{electron}\int_{dyon}.
\end{equation}
It is difficult to calculate this integral in general case hence
we'll do it for the case $M_{d}\gg m$, so we can consider
trajectory of electron as semi circle and trajectory of dyon as a
straight line. Using the dimensional regularization we get
\begin{equation}
\int_{electron}\int_{electron}=
\frac{\pi^{2}}{2}\int_{0}^{1}\frac{\cos\pi(t-\tau)}{1-\cos\pi(t-\tau)}dtd\tau=-\frac{\pi^{2}}{2}.
\end{equation}
Integration for dyon contribution gives
\begin{equation}
\int_{dyon}\int_{dyon}=0,
\end{equation}
and for electron-dyon part
\begin{equation}
\int_{-1}^{1}\int_{-\frac{\pi}{2}}^\frac{\pi}{2}dtd\tau\frac{\cos
t} {(\tau-\sin t)^{2}+\cos^{2} t}
=\frac{\pi}{2}\int_{-\frac{\pi}{2}}^{\frac{\pi}{2}}dt=\frac{\pi^{2}}{2}.
\end{equation}
Finally, collecting all together we find for the case
$M_{d}^{2}<M_{m}^{2}+m^{2}$ the probability of monopole decay is
of the form
\begin{equation}
\begin{array}[c]{rl}
\Gamma\sim&\exp
\biggl(\frac{m^{2}}{eE}\arccos\frac{M_{m}^{2}+m^{2}-M_{d}^{2}}{2mM_{m}}+
\frac{M_{d}^{2}}{eE}\arccos\frac{M_{m}^{2}-m^{2}+M_{d}^{2}}{2M_{d}M_{m}}-
\\ & -\frac{mM_{m}}{eE}\sqrt{1-\left(
\frac{M_{m}^{2}+m^{2}-M_{d}^{2}}{2mM_{m}}\right)^{2}}+
\frac{3e^{2}}{16}\biggr), \label{exp for prob}
\end{array}
\end{equation}
while for the case $M_{d}^{2}>M_{m}^{2}+m^{2}$ similar exponential
factor corresponds to the monoples mass renormalization.
\begin{equation}
\delta m=A\mathrm{e}^{-S_{cl}+S_{int}}.
\end{equation}
For the case $M_{d}^{2}=M_{m}^{2}+m^{2}$, i.e. for BPS case the
negative eigenmode (\ref{specific eigenmode}) becomes zero
eigenmode and the probability for the decay reads as
\begin{equation}
\begin{array}[c]{rl}
\Gamma\sim&\exp \biggl(\frac{\pi m^{2}}{2 eE}+
\frac{(M_{m}+m)^{2}}{eE}\arccos\left(\frac{1}{2}\left(1+\frac{m^{2}}{M_{m}^{2}}\right)^{-\frac{1}{2}}\right)
-\frac{mM_{m}}{eE}+ \frac{3e^{2}}{16}\biggr). \label{exp for prob}
\end{array}
\end{equation}
\section{Conclusion}
In this paper the process of monopole decay in external electric
field was investigated. The probability of monopole decay and
W-boson decay up to the exponential accuracy has been found. It
was shown that the result obtained is in agreement with one for
the probability of pair production. The the Coulomb corrections
were calculated for the process of monopole decay in external
electric field. The determinant was investigated qualitatively,
calculation of full expression for the determinant will be made
elsewhere. It was found that monopole decay doesn't occur for
arbitrary masses of particles involved. The relation between
masses for the decay to exist was found (\ref{condition on
masses}). Note that the well-known process of induced decay of the
false vacuum \cite{selvol} satisfies this condition. It is obvious
that the probability of such processes is exponentially
suppressed, but they could have some applications in cosmology.
The natural generalization of the problem would involve the
temperature effects and the stringy corrections.

The work was supported in part by grant RFBR 04-01-00646 and grant
of support of scientific schools NSh-1999.2003.2

I would like to thank M.B. Voloshin for discussion on the presence
of negative eigenmodes and especially A.S. Gorsky for continuous
attention to the work and for useful and productive discussions.


\newpage
\begin{thebibliography}{10}
\bibitem{sch} J. Schwinger, Phys. Rev. 82, 664 (1951).
\bibitem{aff2} I.K. Affleck, O. Alvarez, N.S. Manton, Nucl. Phys. B197, 509,
(1982).
\bibitem{aff1} I.K. Affleck, N.S. Manton, Nucl. Phys. B194, 38,
(1982).
\bibitem{kobz vol okun} M.B. Voloshin, I. Yu. Kobzarev, L.B.
Okun, Yad. Fiz. 20, 1229 (1974)
\bibitem{Coleman2} C.G. Callan, S. Coleman, Phys. Rev. D16, 1762,
(1977)
\bibitem{Coleman1} S. Coleman, Phys. Rev. D15, 2929 (1977)
\bibitem{gor} A.S. Gorsky, K.A. Saraikin and K.G. Selivanov, Nucl.Phys., B628, (2002),
270-294, hep-th/0110178.
\bibitem{selvol} K. Selivanov and M. Voloshin, ZHETP Lett,42
(1985) 422.
\bibitem{bachporr} C. Bachas and M. Porrati, Phys. Lett. B 296, 77
(1992), hep-th/9209032.
\bibitem{inhom fields} Gerald V. Dunne, C. Schubert "Worldline
instantons and pair production in inhomogeneous fields",
hep-th/0507174
\bibitem{Gor Sel junct} A.S. Gorsky, K.G. Selivanov
"Junction and the Fate of Branes in External Fields", Nucl.Phys.
B571 (2000) 120-134, hep-th/9904041
\bibitem{pioline troost} B. Pioline, J. Troost "Schwinger
pair production in $AdS_{2}$", JHEP 0503 (2005) 043,
hep-th/0501169
\end{thebibliography}
\end{document}